\documentclass[showkeys]{revtex4}
\usepackage[cp866]{inputenc}   % 1
\usepackage[T2A]{fontenc}      % 2
\usepackage{amsmath}           % 5
\usepackage{amssymb}           % 6
\usepackage[dvips]{graphicx}   % 7
\usepackage{euscript}

\usepackage{amssymb}
\usepackage{amsmath}
\usepackage{euscript}
\usepackage[dvips]{graphicx}
\usepackage{epsfig}
\begin{document}

\title{Hamiltonian reduction for the magnetic dynamics
in antiferromagnetic crystals}

\author{D.O.Sinitsyn}

\begin{abstract}

The nonlinear spin dynamics in antiferromagnetic crystals is studied
for the magnetic structures similar to that of hematite.
For the case when only two magnetization vectors are non-zero
and the Hamiltonian has an axial symmetry, a reduction
to a Hamiltonian system with one degree of freedom is performed,
based on the corresponding conservation law. The analysis of the phase portraits
of this system provides tractable analytical and geometric descriptions of the regimes of nonlinear
spin dynamics in the crystal.

\end{abstract}


\date{\today}

\maketitle

\section{Introduction}

The phenomenological method of describing the properties of crystals with magnetic order
is based on attributing classical magnetization vectors $ \vec S_i \ (i=1, \ldots n) $
to the magnetic sublattices of a crystal ($n$ is the number of the sublattices), \cite{Borovik}.
These vectors are subject to exchange interaction between themselves, anisotropic interaction
with the crystal lattice, and, optionally, Zeeman interaction with external magnetic fields, \cite{AM}.
The dynamics of these magnetizations can be described via the corresponding Hamiltonian system.
Generally, the latter is nonlinear and its solving presents substantial difficulties.
In the regimes when the spin vectors are close to their equilibrium positions, the linear
approximation can be used, which allows to obtain the spin wave spectra in a relatively
straightforward manner. However, the general nonlinear dynamics of sublattice magnetizations
is also of considerable interest, e.g. for nonlinear regimes of magnetic resonance.
Hence there is a need for methods of qualitative and quantitative investigating those
dynamics in various situations.

To this end, the analytical tools devised for the study of mechanical Hamiltonian systems can
be employed. In the present paper we apply an approach based on the reduction of a Hamiltonian
system to another Hamiltonian system of lower dimensionality.
The key idea is the considering of a subalgebra of the Poisson algebra of dynamic variables.
If a subalgebra contains the Hamiltonian or, more generally, the Hamiltonian depends on the
elements of the subalgebra and several first integrals, then the elements of the subalgebra
are subject to a new Hamiltonian system with a lower number of phase variables.
It was Poincare who first performed a reduction of a similar type in the three-body problem
(see \cite{Whit}). A reduction based on considering a Poisson subalgebra was applied
to the magnetization dynamics in superfluid $^3He$ B in \cite{Golo},
lowering the phase space dimensionality from 6 to 3.

In the present work we consider the class of Hamiltonians having an axial
symmetry. As a main example we use the case described in the classical
work by Dzyaloshinsky, \cite{Dzyal}, -- the Hamiltonian comprising the
leading terms in the magnetic energy of the four sublattices of
the antiferromagnetic crystal $\alpha$-$Fe_2O_3$.
It belongs to rhombohedral system, having a
third-order symmetry axis, \cite{Dzyal}.
The magnetic energy comprises exchange terms
and several spin-lattice interaction terms, some of them corresponding
to the so-called Dzyaloshinsky-Moriya field leading to week ferromagnetism of $\alpha$-$Fe_2O_3$
(other possibilities include weak additional antiferromagnetism
-- the case of $Cr_2O_3$ in \cite{Dzyal}, and the distortion of magnetic structure
described in \cite{MarTikh}).
In our case, if only second-order terms and the largest fourth-order term are taken
into account (see \cite{Dzyal}, section "Ferromagnetism of $\alpha$-$Fe_2O_3$"),
then the Hamiltonian is invariant with respect to
the rotation about the axis of the crystal. This results into the sum of the
longitude angles of the spins being a cyclic variable, i.e. not
entering the Hamiltonian (for a proper choice of dynamic
variables). This allows to perform a reduction to a Poisson subalgebra
as outlined above, decreasing the number of phase variables by 2 units.
In the case of $\alpha$-$Fe_2O_3$, where only the ferromagnetic vector
and one antiferromagnetic vector are non-zero, this reduction leads
to a system with one degree of freedom, which can be effectively
investigated by means of phase portraits. That provides a detailed
picture of the nonlinear regimes of the spin dynamics in the
given approximation. The same results
hold for the carbonates of Fe, Mn and Co with a difference only in
the values of the parameters of the Hamiltonian.

\section{The Hamiltonian formulation}

We follow the well-known paper \cite{Dzyal} by Dzyaloshinsky for constructing the Hamiltonian system
describing the spin dynamics in $\alpha$-$Fe_2O_3$. The unit cell of the crystal is rhombohedral.
The crystal has 4 magnetic sublattices, and the four corresponding
Fe ions in the unit cell lie on the body diagonal of the rhombohedron -- the third order symmetry axis of the crystal.
The magnetizations of the sublattices are denoted $\vec s_1, \vec s_2, \vec s_3, \vec s_4$.
The Poisson brackets between their components have the usual form of the brackets for angular momentum:
\begin{equation}
\{s_{i \alpha}, s_{i \beta}\} = \varepsilon_{\alpha \beta \gamma} s_{i \gamma}, \quad
\{s_{i \alpha}, s_{j \beta}\} = 0, \ i \ne j,
\label{sbracket}
\end{equation}
where the Greek letters denote the Cartesian coordinates of the vectors
(here and below repeated indices imply summation over 1,2,3).
The degeneracy of these brackets leads to the existence of four Casimir functions
$ s_i^2, \ i=1,\ldots,4$.

Following Dzyaloshinsky, we describe the magnetic ordering in terms of the following four vectors:
$$
\begin{array}{lcl}
\vec m   = \vec s_1 + \vec s_2 + \vec s_3 + \vec s_4,\\
\vec l_1 = \vec s_1 - \vec s_2 - \vec s_3 + \vec s_4,\\
\vec l_2 = \vec s_1 - \vec s_2 + \vec s_3 - \vec s_4,\\
\vec l_3 = \vec s_1 + \vec s_2 - \vec s_3 - \vec s_4.\\
\end{array}
$$
The vector $\vec m$, the total magnetic moment, corresponds to ferromagnetism
(if it is the only non-zero vector, the ordering is ferromagnetic),
the vectors $ \vec l_i $ -- to antiferromagnetism,
each of them describing a particular pattern of antiferromagnetic ordering, \cite{Dzyal}.
The Poisson brackets for these variables, following from (\ref{sbracket}), read:
$$
\{m_{\alpha}, m_{\beta}\} = \varepsilon_{\alpha \beta \gamma} m_{\gamma}, \quad
\{m_{\alpha}, l_{i\beta}\} = \varepsilon_{\alpha \beta \gamma} l_{i \gamma}, \quad
\{l_{i\alpha}, l_{i\beta}\} = \varepsilon_{\alpha \beta \gamma} m_{\gamma}, \quad
\{l_{i\alpha}, l_{j\beta}\} = \varepsilon_{i j k} \, \varepsilon_{\alpha \beta \gamma} \, l_{k \gamma}, \ i \ne j,
$$
At the temperatures close to that of the antiferromagnetic transition the magnetic energy
can be expanded in powers of $\, \vec m, \, \vec l_i $, as their values are small.
Let us use the rectangular coordinate system with the z-axis directed along the axis of the crystal,
the x-axis -- along one of the second-order symmetry axes.

The form of the possible terms in the magnetic energy can be deduced from the analysis of the magnetic symmetry.
A thorough description of this approach is given in \cite{AM}.
As is shown in \cite{Dzyal}, in the case of $\alpha$-$Fe_2O_3$ symmetry restrictions
lead to the following general form of the second-order terms in the magnetic energy of the system:
$$
E = A_1 \vec l_1^2 + A_2 \vec l_2^2 + A_3 \vec l_3^2 + B \vec m^2
+ \alpha_1 l_{1z}^2 + \alpha_2 l_{2z}^2 + \alpha_3 l_{3z}^2 + b\, m_z^2
+ \beta_1 (l_{1x} m_y - l_{1y} m_x) + \beta_2 (l_{2x} l_{3y} - l_{2y} l_{3x}).
$$
Here the first four terms correspond to the exchange interaction, the other terms
-- to the relativistic spin-lattice interaction and the magnetic dipolar interaction.
Experimental results show that in $\alpha$-$Fe_2O_3$ the antiferromagnetic structure
corresponds to the vector $ \vec l_1 $, \cite{Dzyal}, which means that in equilibrium approximately
$ \vec s_1 = - \vec s_2 = - \vec s_3 = \vec s_4 $. There is also weak ferromagnetism
described by the vector $ \vec m $. So, by physical considerations,
the system can be restricted to only these two vectors, which leads to
the Poisson brackets:
\begin{equation}
\{m_{\alpha}, m_{\beta}\} = \varepsilon_{\alpha \beta \gamma} m_{\gamma}, \quad
\{m_{\alpha}, l_{\beta}\} = \varepsilon_{\alpha \beta \gamma} l_{\gamma}, \quad
\{l_{\alpha}, l_{\beta}\} = \varepsilon_{\alpha \beta \gamma} m_{\gamma}, \quad
\label{mlbracket}
\end{equation}
and the Hamiltonian, \cite{Dzyal}:
\begin{equation}
H = \frac{A}2 \, \vec l^2 + \frac{B}2 \, \vec m^2 + \frac{\alpha}2 \, l_z^2 + \frac{b}2 \, m_z^2
+ \beta \, (l_x m_y - l_y m_x) + \frac{C}4 \, \vec l^4,
\label{Ham}
\end{equation}
where $ \vec l = \vec l_1 $, the index 1 is omitted in all positions, and normalizing factors are introduced.
It should be noted that this Hamiltonian has a forth-order term depending on the main magnetic vector $ \vec l = \vec l_1 $.
This model will be the object of our consideration.

\section{Symmetry and reduction}

It is easy to notice that the Hamiltonian (\ref{Ham}) is invariant with respect to the rotation about the z-axis,
i.e. the anisotropy axis.
This leads to $ m_z $ being a first integral. Indeed, calculation shows that $ \dot m_z = \{ m_z, H \} = 0 $.
Moreover, the Poisson algebra (\ref{mlbracket}) has two Casimir functions:
$$
(\vec m + \vec l)^2, \ \ (\vec m - \vec l)^2.
$$
To take advantage of these facts, let us introduce the following variables:
$$
\vec g = \frac{\vec m + \vec l}2, \ \ \vec h = \frac{\vec m - \vec l}2,
$$
which are simply $ \vec g = \vec s_1 + \vec s_4, \ \vec h = \vec s_2 + \vec s_3 $
in terms of the sublattice magnetizations. Their Poisson brackets have the usual form for angular momentum:
$$
\{g_{\alpha}, g_{\beta}\} = \varepsilon_{\alpha \beta \gamma} g_{\gamma}, \quad
\{h_{\alpha}, h_{\beta}\} = \varepsilon_{\alpha \beta \gamma} h_{\gamma}, \quad
\{g_{\alpha}, h_{\beta}\} = 0.
$$

The structure of the Poisson algebra given above admits of a
reduction that substantially simplifies investigating the dynamics of
the system. The key point is to cast the Poisson brackets for vector
dynamical variables in a form that relies on the scalar ones. The
idea is essentially due to K. Pohlmeyer, \cite{Pohl},
who employed it for  studying the algebra of currents
in field theory.  In paper \cite{Golo} the method was used to study
the spin dynamics in the B-phase of superfluid $^3He$ in the regime
of turned off magnetic field.

In our case the reduction is obtained through the following system of variables:
$$
\begin{array}{lcl}
\vec g = (\sqrt{g^2-g_z^2} \, \cos{\varphi_g}, \ \sqrt{g^2-g_z^2} \, \sin{\varphi_g}, g_z), \\
\vec h = (\sqrt{h^2-h_z^2} \, \cos{\varphi_h}, \ \sqrt{h^2-h_z^2} \, \sin{\varphi_h}, h_z).
\end{array}
$$
The new variables have the following sense: $ g $ -- the modulus of
$ \vec g $, $ g_z $ -- its z-component, and $ \varphi_g $ -- its
longitude angle (in xy-plane).

Thus, we take into account the
symmetry of the system and turn to the scalar quantities,  necessary
for the Pohlmeyer reduction.  It is worthwhile to note that a
similar transformation was used in papers  \cite{Sriv1},
\cite{Reichl} for studying chaotic dynamics in two-spin systems.

Using the reverse transformation:
$$
g = \sqrt{g_x^2 + g_y^2 + g_z^2}, \quad \varphi_g = \arctan{\frac{g_y}{g_x}},
$$
we obtain the following Poisson brackets:
$$
\begin{array}{lcl}
\{ \varphi_g, g_z \} = 1, \ \ \{ g, g_z \} = 0, \ \ \{ g, \varphi_g \} = 0, \\
\{ \varphi_h, h_z \} = 1, \ \ \{ h, h_z \} = 0, \ \ \{ h, \varphi_h \} = 0,
\end{array}
$$
all the g--h brackets being zero.

Thus, we have got two advantages. The first one is the lowering of
the phase space dimensionality by two units, as $g$ and $h$ are
Casimir functions, and their values can be fixed. Therefore, we have
a system with two pairs of canonically conjugate variables:
$\varphi_g, g_z$ and $\varphi_h, h_z$. The second one is that we can
easily employ the fact that $ m_z = g_z + h_z $ is a first integral.
Indeed, another formulation of this is that the system is invariant
with respect to the rotation about the z-axis. This means that the
Hamiltonian should depend only on the difference of the angles
$\varphi_g - \varphi_h$, not on their specific values. In fact,
calculation of the Hamiltonian after the above transformations
gives:
$$
\begin{array}{lcl}
\displaystyle
H =
\frac12 \, A \left[-2 \sqrt{g^2-g_z^2} \sqrt{h^2-h_z^2} \, \cos {(\varphi_g-\varphi_h)}-2 g_z h_z+g^2+h^2\right] + \\
+\frac12 \, B \left[2 \sqrt{g^2-g_z^2} \sqrt{h^2-h_z^2} \, \cos {(\varphi_g-\varphi_h)}+2 g_z h_z+g^2+h^2\right] + \\
+\frac12 \, \alpha \, (g_z-h_z)^2 + \frac12 \, b \, (g_z+h_z)^2
-2 \beta \, \sqrt{g^2-g_z^2} \sqrt{h^2-h_z^2} \, \sin{(\varphi_g-\varphi_h)} + \\
+\frac14 \, C \left[-2 \sqrt{g^2-g_z^2} \sqrt{h^2-h_z^2} \cos{(\varphi_g-\varphi_h)}-2 g_z h_z+g^2+h^2\right]^2.
\end{array}
$$
Applying the canonical transformation:
$$
\begin{array}{lcl}
u = \displaystyle \frac{\varphi_g-\varphi_h}{\sqrt2}, \ \ p_u = \frac{g_z - h_z}{\sqrt2}, \\
v = \displaystyle \frac{\varphi_g+\varphi_h}{\sqrt2}, \ \ p_v = \frac{g_z + h_z}{\sqrt2}, \\
\end{array}
$$
we obtain the Hamiltonian:
\begin{equation}
\begin{array}{lcl}
H =

   \frac{1}{2} A \left(-\cos \left(\sqrt{2} u\right) \sqrt{2 g^2-\left(p_u+p_v\right){}^2}
   \sqrt{2 h^2-\left(p_u-p_v\right){}^2}+g^2+h^2+p_u^2-p_v^2\right) \\

   +\frac{1}{2} B \left(\cos \left(\sqrt{2} u\right) \sqrt{2 g^2-\left(p_u+p_v\right){}^2} \sqrt{2
   h^2-\left(p_u-p_v\right){}^2}+g^2+h^2-p_u^2+p_v^2\right) \\

   + \alpha p_u^2 + b p_v^2
   - \beta  \sin \left(\sqrt{2} u\right) \sqrt{2 g^2-\left(p_u+p_v\right){}^2} \sqrt{2 h^2-\left(p_u-p_v\right){}^2} \\

   + \frac{1}{4} C \left(-\cos \left(\sqrt{2} u\right)
   \sqrt{2 g^2-\left(p_u+p_v\right){}^2} \sqrt{2
   h^2-\left(p_u-p_v\right){}^2}+g^2+h^2+p_u^2-p_v^2\right){}^2. \\
\end{array}
\label{HamRed}
\end{equation}

It is easy to see that $v$ does not enter this function. So, as noted above, $p_v$ is a first integral,
and as the variables $u, p_u$ have zero Poisson brackets with it, we conclude that the Hamiltonian \ref{HamRed} defines a
separate Hamiltonian system in variables $u, p_u$ only. In other
words, after fixing the integrals of motion the function
\ref{HamRed} becomes a member of the Poisson subalgebra
of dynamic variables generated by $u, p_u$ and the first integrals, thus defining
Hamiltonian dynamics within this subalgebra.

The study of the dynamics of our system then proceeds as follows.
Firstly,  we study phase pictures  of the reduced system for $u, p_u$.
Secondly, we consider the "lift" of the obtained two-dimensional dynamics to
the six-dimensional space of the initial spin variables.  Results
obtained in this way admit of graphic representation.

\begin{figure}
  \begin{center}
    \includegraphics[width = 400bp]{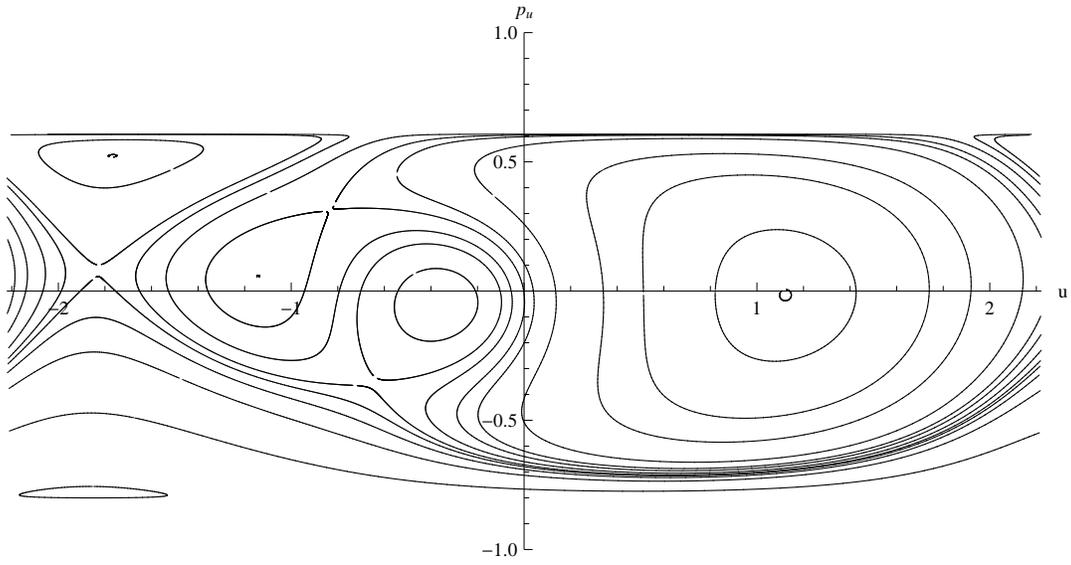}
    \caption{A typical phase portrait of the system for the variables $u, \ p_u = l_z / \sqrt2$.}
    \label{fig1}
  \end{center}
\end{figure}

The dynamical regimes of the reduced system can be studied by means of phase portraits
in the $u\,-\,p_u$ phase plane.
A typical portrait is shown in fig. \ref{fig1}.
It contains several fixed points: centers and saddles, and separatrices, each of which
either connects a pairs of saddles
or forms a loop having the origin and the end in the same saddle.

\begin{figure}
  \begin{center}
    \includegraphics[width = 400bp]{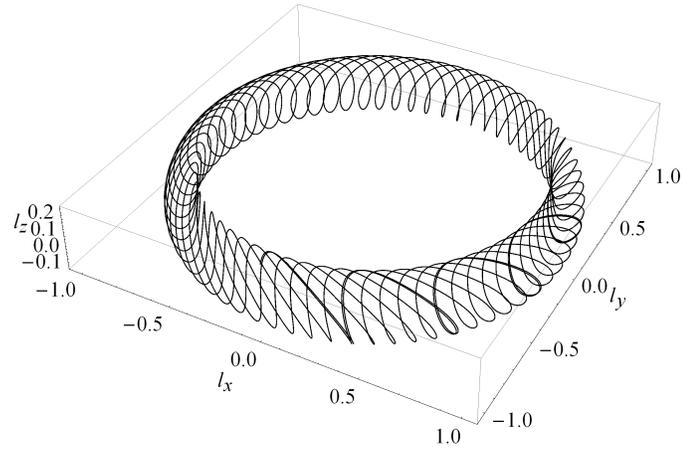}
    \caption{A trajectory surrounding a center (elliptic fixed point) in the phase portrait (fig. \ref{fig1}),
             shown in the space of the coordinates of the antiferromagnetic vector $\vec l$.
             The curve is winding over a 2D-torus.}
    \label{fig2}
  \end{center}
\end{figure}

\begin{figure}
  \begin{center}
    \includegraphics[width = 400bp]{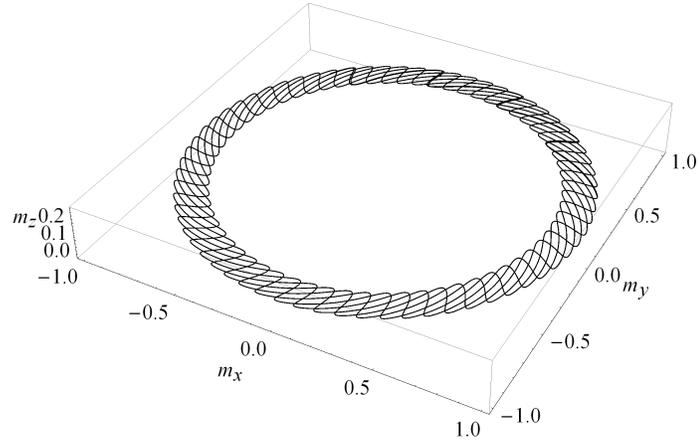}
    \caption{A trajectory surrounding a center (elliptic fixed point) in the phase portrait (fig. \ref{fig1}),
             shown in the space of the coordinates of the ferromagnetic vector $\vec m$.
             The curve is confined to a horizontal plane $m_z = const$.}
    \label{fig3}
  \end{center}
\end{figure}

\begin{figure}
  \begin{center}
    \includegraphics[width = 350bp]{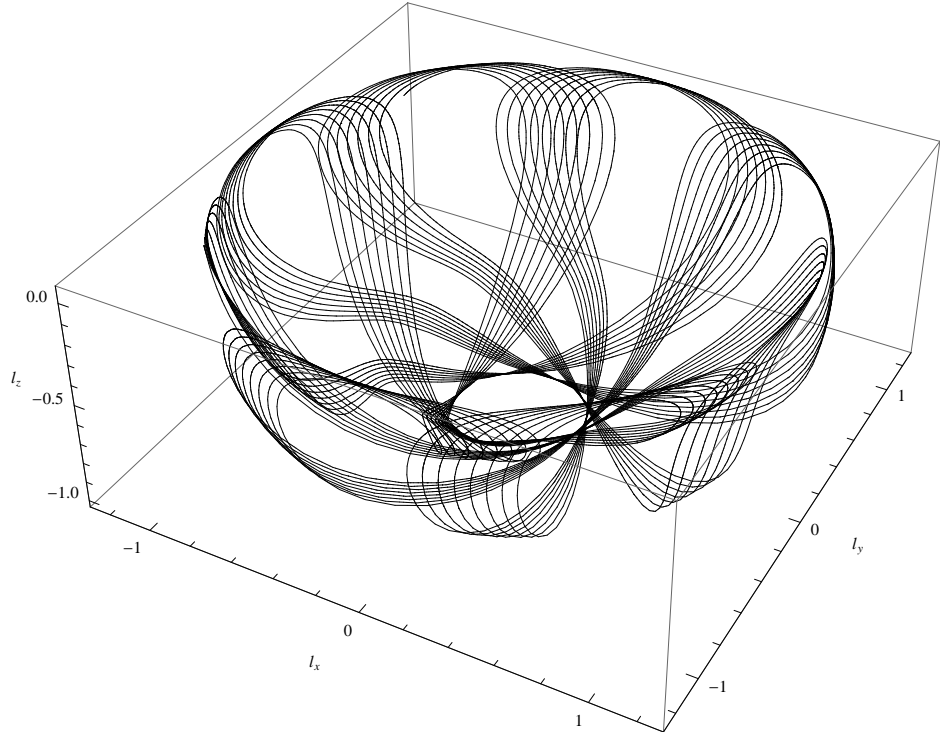}
    \caption{A trajectory close to a separatrix connecting two saddles in the phase portrait (fig. \ref{fig1}),
             shown in the space of the coordinates of the antiferromagnetic vector $\vec l$.
             The vector migrates between two horizontal circles corresponding to the two saddles.}
    \label{fig4}
  \end{center}
\end{figure}

\begin{figure}
  \begin{center}
    \includegraphics[width = 350bp]{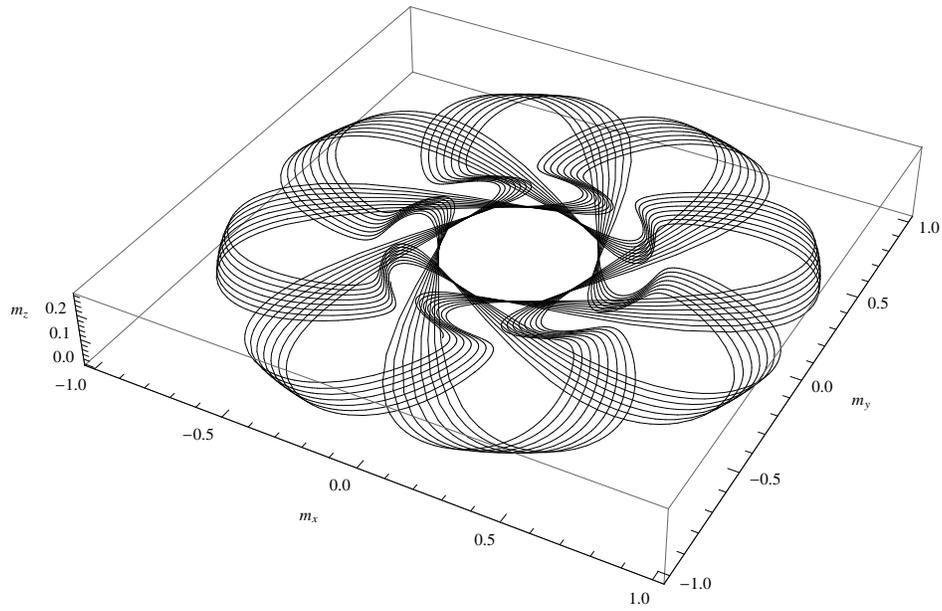}
    \caption{A trajectory close to a separatrix connecting two saddles in the phase portrait (fig. \ref{fig1}),
             shown in the space of the coordinates of the ferromagnetic vector $\vec m$.
             The vector migrates between two circles corresponding to the two saddles.
             The curve is confined to a horizontal plane $m_z = const$.}
    \label{fig5}
  \end{center}
\end{figure}

\begin{figure}
  \begin{center}
    \includegraphics[width = 350bp]{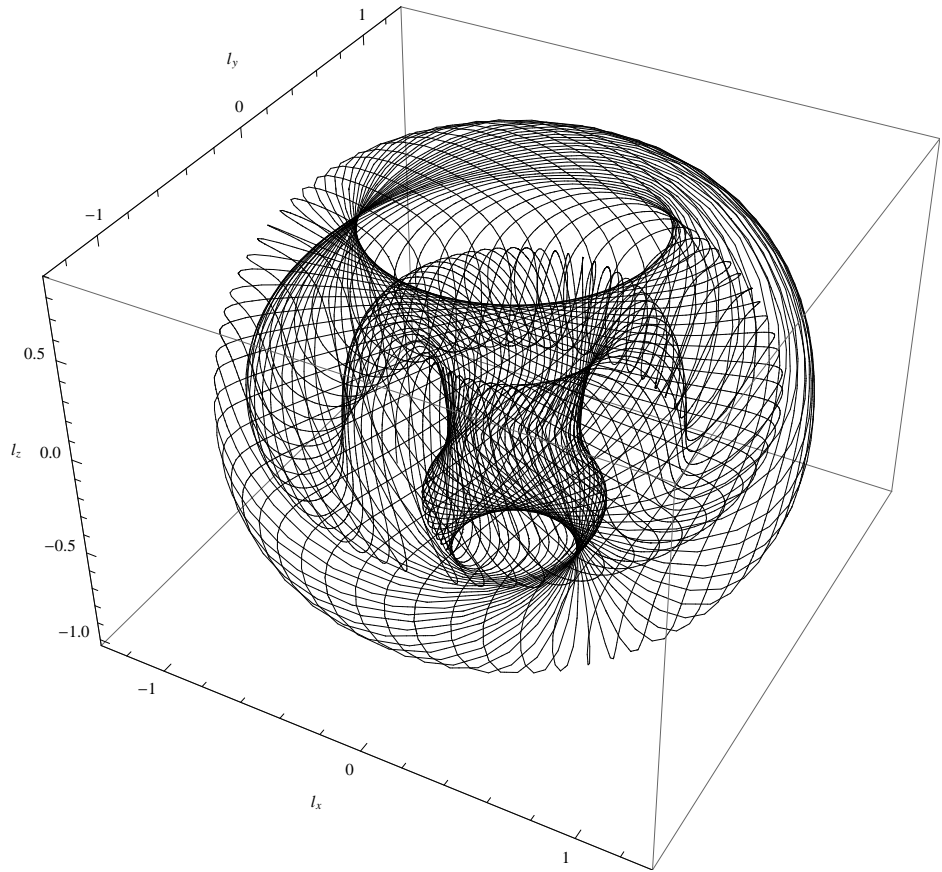}
    \caption{A trajectory taking S-shaped path between several centers and saddles in the phase portrait (fig. \ref{fig1}),
             shown in the space of the coordinates of the antiferromagnetic vector $\vec l$.}
    \label{fig6}
  \end{center}
\end{figure}

\begin{figure}
  \begin{center}
    \includegraphics[width = 350bp]{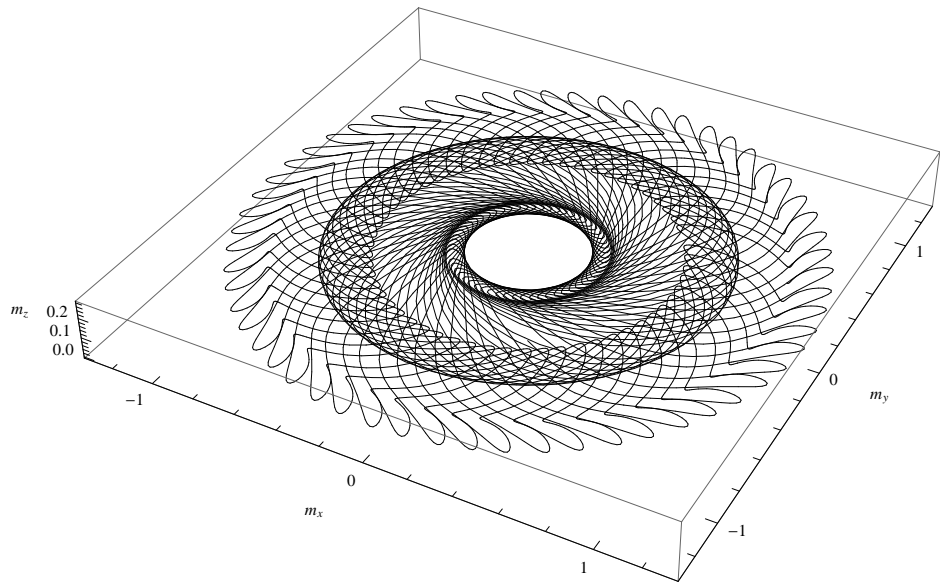}
    \caption{A trajectory taking S-shaped path between several centers and saddles in the phase portrait (fig. \ref{fig1}),
             shown in the space of the coordinates of the ferromagnetic vector $\vec m$.
             The curve is confined to a horizontal plane $m_z = const$.}
    \label{fig7}
  \end{center}
\end{figure}

A typical trajectory going around a center not far from it is shown in figs. \ref{fig2}, \ref{fig3}
in the spaces of the vectors $\vec l$, $\vec m$ respectively;
a trajectory close to a separatrix -- in figs. \ref{fig4}, \ref{fig5}.
It's easy to see that stationary points in the phase portrait correspond
to horizontal circles in the spaces of the magnetic vectors $\vec l$, $\vec m$,
i.e. to circular precession of these vectors.
Other trajectories manifest more complicated behaviour, as can be seen
in figs. \ref{fig6}, \ref{fig7}.

An especially interesting example is a trajectory going very close to a separatrix.
This means that for a long period of time it approaches a saddle point,
manifesting roughly circular precession of magnetizations, but at a certain moment it starts
moving away from that saddle performing some complicated dynamics before reaching
another almost-steady regime of circular precession.

The author is grateful to prof. V.L. Golo for constant attention to this work.

The author acknowledges prof. V.I. Marchenko for useful communications.

The work was supported by the grants RFFI 09-02-00551, 09-03-00779.

\end{document}